\documentclass[dvips,12pt]{article}
\usepackage{graphicx,color}
\newcommand\figcaption{\def\@captype{figure}\caption}
\newcommand\tabcaption{\def\@captype{table}\caption}
\def\al{\alpha}

\def\nb{\nabla}
\def\pa{\partial}
\def\vf{\varphi}
\def\om{\omega}
\def\Om{\Omega}

\def\ga{\gamma}
\def\be{\beta}
\def\dl{\delta}

\def\La{\Lambda}
\def\th{\theta}
\def\sg{\sigma}
\def\Up{\Upsilon}
\def\ck{\check}

\def\nn{\nonumber}
\def\diag{\mbox {diag}}

\title{\Large\bf A Cosmological Model with Dark Spinor Source}
\author{Ying-Qiu Gu\footnote{email: yqgu@fudan.edu.cn}}
\date{\small School of Mathematical Science, Fudan University, Shanghai 200433, China}

\begin{document}
\maketitle
\begin{abstract}
In this paper, we discuss the system of
Friedman-Robertson-Walker(FRW) metric coupling with massive
nonlinear dark spinors in detail, where the thermodynamic movement
of spinors is also taken into account. The results show that, the
nonlinear potential of the spinor field can provide a tiny
negative pressure, which resists the Universe to become singular.
The solution is oscillating in time and closed in space, which
approximately takes the following form
$$
g_{\mu\nu}=\bar R^2(1-\delta\cos t)^2\diag(1,-1,-\sin^2r ,-\sin^2r
\sin^2\theta),
$$
with $\bar R= (1\sim 2)\times 10^{12}$ light year, and
$\delta=0.96\sim 0.99$. The present time is about $t\sim
18^\circ$.


\end{abstract}

\section{Introduction}
Combined with the fluid model, the general relativity usually
gives the cosmological models with singularity. After a series of
singularity theorems, which is based on the conditions such as
positive energy, closed trapped surface, have been proved by
Penrose, Hawking and so on\cite{1,2,3,4}, it seems that the
singularity has been quite naturally accepted as an objective
existence. However the intrinsic singularities are after all
abnormal and inconceivable. The fluid model is just a simplified
description of matter, so one is unnecessary to take it too
serious if the intrinsic essences of the spacetime and matter are
involved.

The global solutions of Einstein's equation are always
interesting\cite{6}-\cite{gu1}. The Lemaitre's Phoenix picture, a
regular and oscillating cosmological model is reconsidered in
\cite{13,14}, it can give a more natural explanation for some
famous cosmological puzzles such as the flatness and horizon
problems. However this model is obtained by hypothesis rather than
by solution of Einstein equation. Recently the coupling system of
spinor field with spacetime has been studied by several authors.
The principal motivation of the works was to find out the regular
solutions of the corresponding field equations\cite{5}-\cite{17}
or explain the accelerating expansion of the present
Universe\cite{18}-\cite{21}. These works indeed obtained
singularity-free solutions, and showed that the nonlinear spinor
field results in a rapid expansion of the Universe.

In what follows, we also investigate the cosmological model with
nonlinear spinor fields, which is derived from a general framework
of field theory\cite{22}-\cite{gu2}. Different from the treatment
of \cite{15}-\cite{21}, in which the authors introduced only one
spinor flied independent of spatial coordinates, and suffered from
the energy momentum tensor incompatible with the high symmetry of
the Friedman-Robertson-Walker(FRW) metric. In our point of view,
one spinor field just describes one particle and the high symmetry
is a result in mean sense. Then all problems can be easily solved,
and the result is quite natural and reasonable.

\section{The Basic Equations}
\setcounter{equation}{0}

A large number of empirical data shows that our Universe is highly
homogeneous and isotropic. So the Universe can be described by FRW
metric in mean sense, the corresponding line element is given by
\begin{equation}
ds^2=d\tau^2-a^2(\tau)\left(\frac {d
r^2}{1-Kr^2}+r^2(d\th^2+\sin^2\th d\vf^2)\right), \label{2.1}
\end{equation}
where $K=1,~ 0$ and $-1$ correspond to the closed, flat and open
Universe respectively. However this form of FRW metric is not
convenient for analysis, and the solution is difficult to
expressed by elementary functions. so in this paper we adopt the
conformal coordinate system\cite{14}. The corresponding metric
becomes
\begin{equation}
g_{\mu\nu}=a^2(t)\diag\left(1,-1,-f^2(r),-f^2(r)\sin^2\th\right),
\label{2.2}
\end{equation}
where $d\tau =a(t) dt$,
\begin{equation} f=\left \{ \begin{array}{ll}
  \sin r  & {\rm if} \quad  K=1,\\
   r   & {\rm if} \quad  K=0,\\
  \sinh r  & {\rm if} \quad  K=-1.
\end{array} \right. \label{2.3}
\end{equation}

The Lagrangian of the gravitational field is generally given
by\cite{25}
\begin{equation}
{\cal L}_g=\frac 1 {16\pi G}(R-2\La), \label{2.4}
\end{equation}
where $R$ is the scalar curvature, $\La$ the cosmological factor.
For the metric (\ref{2.2}), we have the scalar curvature and
Einstein tensor $G_{\mu\nu}$ as follows
\begin{eqnarray}
R&=&6\frac{a''+Ka}{a^3}, \label{2.5.1} \\
G_{00}&=&-3\left(\frac {a'^2}{a^2}+K\right),\label{2.5.2}\\
G_{11}&=&2\frac {a''}{a}-\frac {a'^2}{a^2}+K,\label{2.5.3}\\
G_{22}&=&G_{11}f^2,~~G_{33}=G_{11}f^2\sin^2\th,
\label{2.5.4}\end{eqnarray} where $a'=\frac {d}{dt}a,~f'=\frac
{d}{dr}f$.

Now we construct the Lagrangian of matter. Denote the Pauli
matrices by
\begin{equation}
 {\vec\sg}=(\sg^{k})= \left \{\pmatrix{
 0 & 1 \cr 1 & 0},\pmatrix {
 0 & -i \cr i & 0},\pmatrix{
 1 & 0 \cr 0 & -1}
 \right\}.\label{2.6}\end{equation}
Define $4\times4$ Hermitian matrices by
\begin{equation}\al^\mu=\left\{\left ( \begin{array}{ll} I & ~0 \\
0 & I \end{array} \right),\left (\begin{array}{ll} 0 & \vec\sg \\
\vec\sg & 0 \end{array} \right)\right\},
~ \be=\left (\begin{array}{ll} 0 & -iI \\
iI & ~~0 \end{array} \right),~ \ga =\left ( \begin{array}{ll} I & ~0 \\
0 & -I \end{array} \right).\label{2.7}
\end{equation}
Instead of Dirac matrices $\ga^\mu$, we adopt the above Hermitian
matrices (\ref{2.7}) for the convenience of calculation.
Obviously, the most part of matter should be described by spinors.
For some reasons, the following Lagrangian is the most natural
candidate for matter fields, especially for dark matter. In flat
spacetime the Lagrangian is given by
\begin{equation}
{\cal L} =\sum_{k}\phi_k^+(\al^\mu  i\pa_\mu-\mu_k
\ga)\phi_k+F(\ck\al^\mu_k,\ck\be_k,\ck\ga_k),
\label{2.8}\end{equation} where we assign $\phi_k$ to the $k$-th
dark spinor $S_k$, $\mu_k>0$ is constant mass, which takes one
value for the same kind particles,
\begin{equation}
\ck\al^\mu_k=\phi_k^+\al^\mu\phi_k,\quad
\ck\be_k=\phi_k^+\be\phi_k, \quad \ck\ga_k=\phi_k^+\ga\phi_k,
\label{2.8.1}\end{equation} $F$ is the nonlinear coupling term,
which is small in value ($|F|\ll \mu_k$), but decisive for the
structure of matter\cite{29},\cite{27}-\cite{gu3} and the
Universe.

Considering that all the particle-like solutions of the nonlinear
spinor equation have a mean diameter less than $10^2$ Compton wave
lengthes, and all $|\phi_k|$ decay exponentially with respect to
the distance from the center\cite{29,30,gu3}, so we can neglect
the nonlinear interaction terms among spinors, but keep the self
coupling terms only. By the Pauli-Fierz identities
$\ck\al^\mu\ck\al_\mu=\ck\be^2+\ck\ga^2$\cite{38} and the
stability of the fields, $F$ should take the following form
\begin{equation}F=\sum_k V_k, \qquad V_k \equiv V(\ck\ga_k),\label{2.9}\end{equation}
where the self-coupling potential $V$ is a differentiable, even
and concave function satisfying
\begin{equation}V(0)=0,\quad V'(x)x-V(x)>0. \label{2.10}\end{equation}

The covariant form of spinor equation in curved spacetime have
studied by many authors\cite{34}-\cite{gu4}. The generalized form
of (\ref{2.8}) with diagonal metric is given by
\begin{equation}
{\cal L}_m=\sum_k \left( \Re \left( \phi^+_k \varrho^\mu i
\pa_\mu\phi_k\right)-\mu_k\ck \ga_k+V_k\right), \label{2.11}
\end{equation}
in where
\begin{equation}
\varrho^\mu=\left(\frac {\al^0}{a},~\frac {\al^1}{a},~\frac
{\al^2}{af},~\frac {\al^3}{af\sin\th}\right).
\end{equation}
The variation of (\ref{2.11}) with respect to $\phi_k^+$ gives
dynamic equation for $S_k$\cite{32,gu4}
\begin{equation}
\varrho^\mu i\nb_\mu\phi_k=(\mu_k-V'_k)\ga \phi_k, \quad (\forall
k) \label{2.12}
\end{equation}
where $\nb_\mu=(\pa_\mu+\Up_\mu)$, $\Up_\mu$ is the spinor
connection
\begin{equation}
\Up_\mu=\left(\frac{3a'}{2a},~\frac {f'} f ,~\frac 1 2 \cot\th,~0
\right). \label{2.12.1}
\end{equation}

Coupling (\ref{2.4}) with (\ref{2.11}), we get the total
Lagrangian of the Universe with dark spinors
\begin{equation}
{\cal L}=\frac 1 {16\pi G}(R-2\La)+\sum_k \left(\Re \left(
\phi^+_k \varrho^\mu i \pa_\mu\phi_k\right)-\mu_k\ck
\ga_k+V_k\right). \label{2.13} \end{equation} Generally the
variation (\ref{2.13}) with respect to $g_{\mu\nu}$ gives the
Einstein equation
\begin{equation}G^{\mu\nu}+\La g^{\mu\nu}+8\pi G T^{\mu\nu}=0.\label{2.14}\end{equation}
In the spacetime with orthogonal coordinates, the energy momentum
tensor $T^{\mu\nu}$ is relatively simple\cite{31,gu4}
\begin{eqnarray}
T^{\mu\nu}&=&\sum_k\left(\frac 1 2 \Re \left(\phi^+_k(\varrho^\mu
i\pa^\nu+\varrho^\nu
i\pa^\mu)\phi_k\right)-{\cal L}_m g^{\mu\nu}\right)\nn\\
&=&\sum_k\left(\frac 1 2 \Re \left(\phi^+_k(\varrho^\mu
i\nb^\nu+\varrho^\nu
i\nb^\mu)\phi_k\right)+(V'_k\ck\ga_k-V_k)g^{\mu\nu}\right).
\label{2.15}\end{eqnarray} In (\ref{2.15}) we substituted the
dynamic equation (\ref{2.12}) into ${\cal L}_m$.

For the spinor at particle state\cite{gu2,gu5}, we have the
following classical approximation
\begin{eqnarray}\phi^+_k\varrho^\mu\phi_k &\to& u^\mu_k\sqrt{1-v^2_k}\dl(\vec x-\vec X_k),\\
i\nb^\mu\phi_k&\to& p^\mu_k\phi_k=m_k u^\mu\phi_k
\label{2.16}\end{eqnarray} where $\vec X_k$ is the central
coordinate of $S_k$, $u^\mu_k$ is its 4-vector speed. Then the
classical approximation of (\ref{2.15}) provides the energy
momentum tensor of the ideal gas with a tiny negative pressure
term
\begin{equation}T^{\mu\nu}=\sum_k(m_ku^\mu_ku^\nu_k+W_k
g^{\mu\nu})\sqrt{1-v^2_k}\dl^3(\vec x-\vec
X_k),\label{2.17}\end{equation} where
$W_k=\int_{R^3}(V'_k\ck\ga_k-V_k)d^3 \bar x>0$ is the proper
energy of $S_k$ contributed by $V_k$, which acts as negative
pressure. The details of the classical approximation and local
Lorentz transformation see \cite{gu5,gu6}. The numerical results
show that $0<W_k\ll m_k$, but it keep a spinor to be an
independent particle\cite{29,30,gu3}. In what follows we see how
this wizardly tiny term resists the Universe from singularity.

\section{Solution to the Equation}
\setcounter{equation}{0}

The complete Einstein equation (\ref{2.14}) is an overdetermined
system under the assumption of FRW metric, because the FRW metric
only holds in mean sense. For an overdetermined system we should
appeal to the variation principle. The action corresponding to
(\ref{2.13}) is given by
\begin{equation}
{\textbf{I}}=\int dt\int_\Om {\cal L}a^4
d\Om,\label{3.1}\end{equation} where $d\Om=f^2\sin\th drd\th d\vf$
is the angular volume independent of $t$. Variation (\ref{3.1})
with respect to $a$, we get Euler equation as follows
\begin{equation}
\frac {\left(3a''+3Ka-2\La a^3\right)\Om} {4\pi G}+\sum_{X_k\in
\Om}\int_\Om \left(3\Re \left( \phi^+_k \varrho^\mu i
\nb_\mu\phi_k\right)-4\mu_k\ck \ga_k+4V_k\right)a^3
d\Om=0,\label{3.2}
\end{equation}
where $\Om$ is any angular volume large enough to contain
sufficient particles, such that the statistical principle is
valid. Substituting (\ref{2.12}) into (\ref{3.2}) we get
\begin{equation}
a''+Ka-\frac 2 3\La a^3=\frac{4\pi G}{3\Om}\sum_{X_k\in
\Om}\int_\Om \left(\mu_k\ck\ga_k+3V'_k\ck
\ga_k-4V_k\right)a^3d\Om.\label{3.3}
\end{equation}
Define the proper mass of $S_k$ by
\begin{equation}m_k=\int_{R^3}
\left(\mu_k\ck\ga_k+3V'_k\ck \ga_k-4V_k\right)d^3 \bar
x_k,\label{3.4}
\end{equation}
where $\bar x_k$ is the local Cartesian coordinates of the central
coordinate system of $S_k$. Obviously $m_k$ is a constant
independent of $t$. Substituting it into (\ref{3.3}) and making
local Lorentz transformation\cite{gu5}, we get
\begin{equation}
a''+Ka-\frac 2 3\La a^3=\frac{4\pi G}{3\Om}\sum_{X_k\in
\Om}m_k\sqrt{1-v^2_k}.\label{3.5}
\end{equation}
The drifting speed $v_k$ depends on $a(t)$, solving the geodesic
of $S_k$ we have\cite{33}
\begin{equation}
v_k=\frac {b_k}{\sqrt{a^2+b_k^2}},~~~{\rm
or}~~~\sqrt{1-v_k^2}=\frac a
{\sqrt{a^2+b_k^2}},\label{3.6}\end{equation} where $b_k$ is a
constant determined by initial speed. Substituting (\ref{3.6})
into (\ref{3.5}), we get
\begin{equation}
a''+Ka-\frac 2 3\La a^3=\frac{4\pi G}{3\Om}\sum_{X_k\in \Om}\frac
{m_k a} {\sqrt{a^2+b_k^2}}.\label{3.7}
\end{equation}
Multiply (\ref{3.7}) by $a'$ and integrate it, again by
(\ref{3.6}) we have
\begin{equation}
a'^2+Ka^2-\frac 1 3\La a^4=\frac{8\pi G}{3\Om}\sum_{X_k\in
\Om}\frac {m_k a} {\sqrt{1-v_k^2}}+C_0.\label{3.8}
\end{equation}

Now we examine the meaning of $C_0$. By (\ref{3.8}),
(\ref{2.5.1})-(\ref{2.5.4}) and (\ref{2.15}) we have
\begin{equation}\begin{array}{lll}
C_0&=&a'^2+Ka^2-\frac 1 3\La a^4-\frac{8\pi G}{3\Om}\sum_{X_k\in
\Om}\frac {m_k a} {\sqrt{1-v_k^2}}\\
&=&-\frac 1 3 a^2 G_{00}-\frac 1 3\La a^4-\frac{8\pi
G}{3\Om}\sum_{X_k\in
\Om}\frac {m_k a} {\sqrt{1-v_k^2}}\\
&\approx&\frac{8\pi G}{3} \left(a^4 T^0_0-\frac 1 \Om\sum_{X_k\in
\Om}\frac {m_k a} {\sqrt{1-v_k^2}}\right).\end{array}\label{3.9}
\end{equation}
The equivalent pressure for stationary $S_k$ is given by\cite{gu3}
\begin{equation}P_k\equiv \frac 1 3
(T^0_0-T^\mu_\mu)=\frac 1 3 (\phi_k^+\al^0
i\bar\pa_0\phi_k-\mu_k\ck\ga_k-2V'_k\ck\ga_k+3V_k),\end{equation}
where $\bar x_k$ stands for the central coordinate system of
$S_k$. The numerical results show that for ground state
\begin{equation}
\int_{R^3}P_k d^3\bar x_k<0.
\end{equation}
In statistical sense, we have $$\begin{array}{lll}
M_k&\equiv& \int_\Om\Re  \left(\phi_k^+\varrho^0 i\nb_0\phi_k\right) a^3 d\Om\\
&=&\int_{R^3} \frac 1{\sqrt{1-v_k^2}}({\phi_k^+\al^0
i\bar\pa_0\phi_k-v^2_k\phi_k^+\al^1 i\bar\pa_1\phi_k})d^3\bar x_k\\
&=&\int_{R^3} \frac1{\sqrt{1-v_k^2}} \left((1+\frac1 3
v^2_k){\phi_k^+\al^0 i\bar\pa_0\phi_k-\frac 1 3
v^2_k\phi_k^+\al^\mu
i\bar\pa_\mu\phi_k}\right)d^3\bar x_k\\
&=&\int_{R^3} \frac 1 {\sqrt{1-v_k^2}} \left((1+\frac1 3
v^2_k)(3P_k+\mu_k\ck\ga_k+2V'_k\ck\ga_k-3V_k)-\frac 1 3
v^2_k(\mu_k-V'_k)\ck\ga_k\right)d^3\bar x_k\\
&=&\int_{R^3}\frac 1
{\sqrt{1-v_k^2}}\left[(3+v^2_k)P_k+\mu_k\ck\ga_k+2V'_k\ck\ga_k-3V_k+v^2_k(V'_k\ck\ga_k-V_k)
 \right]d^3\bar x_k,\end{array}$$
where we also made local Lorentz transformation for the integrals
of $S_k$. Then we get
\begin{equation}\begin{array}{lll}
T^0_0&=&\frac 1 {a^3\Om} \int_\Om
\sum_k\left(\Re \left(\phi_k^+\varrho^0 i\nb_0\phi_k\right)+V'_k\ck\ga-V_k\right)a^3d\Om\\
&=&\frac 1 {a^3\Om}  \sum_{X_k\in\Om}
\left(M_k+\int_{R^3}(V'_k\ck\ga_k -
V_k)\sqrt{1-v_k^2} d^3\bar x_k\right)\\
&=&\frac 1 {a^3\Om} \sum_{X_k\in\Om} \int_{R^3} \frac1
{\sqrt{1-v_k^2}}\left[(3+v^2_k)P_k+\mu_k\ck\ga_k+3V'_k\ck\ga_k-4V_k\right]d^3\bar
x_k.\end{array}\label{3.10}\end{equation} Substituting (\ref{3.4})
and (\ref{3.10}) into (\ref{3.9}) we get
\begin{equation}
C_0\approx \frac{8\pi G a}{3\Om} \sum_{X_k\in \Om} \frac {3+v_k^2}
{\sqrt{1-v_k^2}}\int_{R^3} P_k d^3 \bar x_k<0. \label{3.11}
\end{equation}
So $C_0$ is a negative quantity related to the nonlinear potential
$V$. Of course (\ref{3.11}) can not be used as definition in
(\ref{3.8}), because $G^0_0$ and $T^0_0$ are ``overdetermined
functions" with respect to FRW metric.

The kinetic energy of $S_k$ is calculated by
\begin{eqnarray}{\textbf{K}}_k=\frac
{m_k}{\sqrt{1-v_k^2}}-{m_k},\label{3.12.1} \end{eqnarray} Although
the spinors are fermions, but if the mean distance among them is
much larger than their mean radiuses, then they are all at
particle states and the effects of exclusion principle actually
vanish. So the kinetic energy of such spinors should satisfy the
classical distribution, rather than Fermi-Dirac statistics.
Considering the case of relativity, we have the following
Maxwell-like distribution
\begin{equation}
f(\textbf{K})d \textbf{K} =\sqrt{\frac
{4{\textbf{K}}}{\pi(kT)^3}}\exp\left(-\frac
{\textbf{K}}{kT}\right)d{\textbf{K}}.\label{3.12.2} \end{equation}
Substituting (\ref{3.12.1}) and (\ref{3.12.2}) into (\ref{3.8}),
we get
\begin{eqnarray}
a'^2+Ka^2-\frac 1 3\La a^4 &=& \frac{8\pi G a}{3\Om}\sum_{X_k\in
\Om} \int^\infty_0\left({\textbf{K}}+{m_k}\right)f(\textbf{K})d
\textbf{K} +C_0 \nn \\
&=& \frac{8\pi G a}{3\Om}\sum_{X_k\in \Om} \left( \frac 3 2 kT +
m_k \right)+C_0.\label{3.12}
\end{eqnarray}
In \cite{33}  we derived the relation $T(a)$ as follows
\begin{equation}
\frac 1 2 kT=\frac{\bar m
b^2}{5a(a+\sqrt{a^2+b^2})},\label{3.13}\end{equation} where $\bar
m$ is mean mass of all particles. $b$ is a constant determined by
the initial data
\begin{equation}
\frac b {a_0}=\sqrt{\frac {5kT_0}{\bar m}\left(1+\frac
{5kT_0}{4\bar m}\right)}\ll 1,\quad(1eV \sim 10^4
K).\label{3.14}\end{equation} Substituting (\ref{3.13}) into
(\ref{3.12}) we finally get
\begin{equation}
a'^2+Ka^2-\frac 1 3\La a^4 = 2\bar R \left( \frac
{3b^2}{5(a+\sqrt{a^2+b^2})} +a \right)+C_0,\label{3.15}
\end{equation}
where \begin{equation}\bar R=\frac {4\pi G} {3\Om} \sum_{X_k\in
\Om}m_k\end{equation} is the radius scale of the Universe, which
is independent of $a$.

In \cite{gu7}, the influences of the constants in (\ref{3.15}) on
the solution were analyzed in detail. The main results are that:
(R1). $b\ll a_0$ mainly influences the behavior as $a\to 0$, it
has little influence on the present Universe. (R2). $\La$ mainly
influences the behavior as $a\ge \bar R$, it seems to be
superfluous term without any effective use,  so $\La=0$ is the
most natural choice. (R3). Eq.(\ref{3.15}) has singularity-free
solution as long as $C_0<-\frac 6 5 \bar R b$.

According to these results, we set $\La=b=0$ for the following
computation, the detailed computation will be given elsewhere.
Then (\ref{3.15}) becomes
\begin{equation}
a'^2+Ka^2= 2\bar Ra+C_0.\label{3.16}
\end{equation}
Solving (\ref{3.16}), we get the singularity-free solutions as
follows
\begin{equation} a=\left \{ \begin{array}{llll}
  \bar R (1-\dl \cos t ),& (C_0=-\bar R^2(1-\dl^2),& 0<\dl<1), & {\rm if~~} f=\sin r ,\\
  \bar R(\dl+\frac 1 2    t^2),& (C_0=-2\bar R^2\dl,& \dl>0),   & {\rm if~~} f=r ,\\
  \bar R(\dl\cosh t -1),& (C_0=-\bar R^2(\dl^2-1),& \dl>1), & {\rm if~~} f=\sinh r .
\end{array} \right. \label{3.17}
\end{equation}
Calculating the energy-momentum tensor $T_{\mu\nu}$ by $8\pi G
T_{\mu\nu}=-G_{\mu\nu}$, we have
\begin{eqnarray}
&&T_{00}=\frac {3(1-2\dl\cos t +\dl^2)}{8\pi G(1-\dl\cos t )^2},~~
T_{11}=\frac{\dl^2-1} {8\pi G(1-\dl\cos t )^2},~~~{\rm
if}~f=\sin r , \label{3.18} \\
&&T_{00}=\frac {3   t^2}{8\pi G(\dl+\frac 1 2 t^2)^2},\qquad\quad
T_{11}=\frac{-2} {8\pi G(\dl+\frac 1 2 t^2)^2},\qquad{\rm
if}~f=r , \label{3.19}\\
&&T_{00}=\frac {3(2\dl\cosh t-1-\dl^2)}{8\pi G(1-\dl\cosh t )^2},~
T_{11}=\frac{1-\dl^2} {8\pi G(1-\dl\cosh t )^2},~{\rm if}~f=\sinh
r. \label{3.20}
\end{eqnarray}
From the above equations we learn that, only for the closed case
the energy $T_{00}$ is always positive, so only this solution is
meaningful in physics. Moreover we find $T_{11}<0$ for all cases.
$T_{11}<0$ is inconceivable for perfect fluid model, but from
(\ref{2.15}) and (\ref{2.17}), we learn that it is normal for
nonlinear spinor field.
\section{Sketch of the Universe}
\setcounter{equation}{0}

In this section we associate the closed case in (\ref{3.17}) with
some observational data and give a sketch of the Universe. The
detailed analysis will be given elsewhere.  Define the
cosmological time by $\tau$
\begin{equation}
\tau=\int_0^ta(t) dt =\bar R( t -\dl\sin t ). \label{4.1}
\end{equation} Then the age of the Universe ${\cal T}$ reads
\begin{equation}
{\cal T}=\bar R( t_a-\dl \sin t_a),\label{4.3} \end{equation}
where $ t_a$ stands for the age in angular coordinate system. The
Hubble's parameter is defined by
\begin{equation}
H=\frac{da}{ad\tau}= \frac {a'}{a^2}=\frac {\dl \sin t_a}{\bar R
(1-\dl\cos t_a)^2}. \label{4.5}
\end{equation}
The mass density $\rho$ and the pressure $P$ in usual sense are
defined by
\begin{eqnarray}
\rho&=&T^0_0=\frac 3 {8\pi G\bar R^2}\cdot\frac {1-2\dl\cos
t_a+\dl^2}{(1-\dl\cos t_a)^4}. \label{4.8} \\
P&=&-T^1_1=\frac {-1} {8\pi G \bar R^2}\cdot\frac{1-\dl^2}
{(1-\dl\cos t_a )^4}. \label{4.8.1}
\end{eqnarray}
Denote the critical density $\rho_c=\frac 3{8\pi G}H^2\sim 8\times
10^{-30}({\rm g/cm}^3)$, we have
\begin{equation}
\Om_{\rm tot}=\frac{\rho}{\rho_c}=1+\frac{a^2}{a'^2}=\frac
{1-2\dl\cos t_a+\dl^2}{(\dl\sin t_a)^2}. \label{4.9}
\end{equation}
The equation of state is defined by
\begin{equation}
w=\frac{P}{\rho}=- \frac{1-\dl^2}{3(1-2\dl\cos t_a+\dl^2)}.
\label{4.9.1}
\end{equation}
By (\ref{4.9.1}) we find $w(t_a)<0$ is an increasing function for
$0\le t_a<\pi$, and
$$w\to -\frac {1+\dl}{3(1-\dl)}, \quad {\mbox{as}}\quad t_a\to
0.$$ So $w$ of spinors is a function crossing the value $-1$.

There are serval authoritative empirical
data\cite{SNeIa1}-\cite{Padma}. We take the data in \cite{SDSS1}
as basic parameters for computation.
\begin{equation}
{\cal T}=14.2({\mbox{Gyr}}), \quad  \Om_{\rm tot}= 1.02\sim
1.08,\quad \om=-1\sim -0.2. \label{4.19}
\end{equation}
By computation we get
\begin{equation}
t_a=12^\circ\sim 22^\circ,~~\dl=0.96\sim0.99, ~~\bar R=(1.0\sim
2.0)\times 10^{12}{\rm yr}. \label{4.7}
\end{equation}
By $ t_a$ we learn that the Universe is very young. The radius of
the present Universe is about $\pi \bar R (1-\dl \cos(t_a)) \sim
300$Gyr. The period for a photon to travel a cycle in the Universe
is just the life period of the Universe herself, so we can never
observe the ghost of a galaxy at any time. However, by the
functional digraphes of $H(t)$, $\Om_{\rm tot}$ and $w(t)$, we
find they are rapidly varying functions of $t$ of the past, then
consequently , the observational data strongly depend on the
distances of the galaxies. So the parameters provided here could
only be used as rough scales, the more values should be obtained
dynamically under the aid and constraint of specific cosmological
models.

From (\ref{3.17}) we learn that, instead of the initial
singularity there is a cradle of rebirth $t=0$. At this time, the
drifting speed of the particles or equivalently the cosmic
temperature takes the maximum, and the volume of the Universe
takes the minimum. This situation will result in violent collision
among galaxies and particles. Such collisions will recover the
vitality of the dead stars and combined particles. Just like the
rebirth of the Phoenix, the Universe gets her rebirth via the
destructive high temperature.

Of course the above analysis are based on the idealized  model of
particles. What is important for this paper is that, it reveals
the solutions of the Einstein's equation are sensitively related
to the properties of the source. The interactional mechanism
between the Universe and fundamental particles has been widely
investigated by the group of the Center for Cosmoparticle
Physics\cite{MYK1}-\cite{MYK13}, and their methods and results
will be important for further study on detailed interaction.

\section*{Acknowledgments}
The author is grateful to his supervisor Prof. Ta-Tsien Li for his
encouragement and guidance. Thanks to Prof. Maxim Yu. Khlopov for
some meaningful discussions.

\end{document}